\documentclass[fleqn,usenatbib]{mnras}

\usepackage[T1]{fontenc}

\DeclareRobustCommand{\VAN}[3]{#2}
\let\VANthebibliography\thebibliography
\def\thebibliography{\DeclareRobustCommand{\VAN}[3]{##3}\VANthebibliography}

\usepackage{graphicx}	
\usepackage{amsmath}	
\usepackage{amssymb}	
\usepackage{newtxtext,newtxmath}

\title[Search for movement of GRB
jets using Swift data]{Statistical search for angular non-stationarities of long gamma-ray burst jets using Swift data}

\author[Budai, Borgulya, Dawes, Szeifert and Raffai]{
A. Budai,$^{1, 2}$\thanks{E-mail: arandras@staff.elte.hu}
P. Raffai,$^{1, 2, 3}$
B. Borgulya,$^{4}$
B. A. Dawes,$^{5}$
and G. Szeifert$^{2}$
\\
$^{1}$E\"{o}tv\"{o}s Lor\'{a}nd University, Institute of Physics, 1117 Budapest, Hungary\\
$^{2}$ELTE Gravitational-wave and Cosmology Research Group, 1117 Budapest, Hungary\\
$^{3}$MTA-ELTE Extragalactic Astrophysics Research Group, 1117 Budapest, Hungary\\
$^{4}$University of Edinburgh, School of Physics and Astronomy, EH9 3JZ, Edinburgh, UK\\
$^{5}$Columbia University, Department of Physics, New York, NY 10027, USA\\
}
\date{Accepted XXX. Received YYY; in original form ZZZ}

\pubyear{2021}

\begin{document}
\label{firstpage}
\pagerange{\pageref{firstpage}--\pageref{lastpage}}
\maketitle

\begin{abstract}
In a previous article we argued that angular non-stationarities of gamma-ray burst (GRB) jets can result in a statistical connection between the angle values deduced from jet break times and the variabilities of prompt light curves. The connection should be an anti-correlation if luminosity densities of jets follow a power-law or a uniform profile, and a correlation if they have a Gaussian profile. In this follow-up paper, we search for the connection by measuring Spearman's rank correlation coefficient in a sample of 19 long GRBs observed by the Swift satellite. Using 16 of the GRBs with well-defined angle measurements, we find $\rho  = -0.38_{-0.1}^{+0.1}$ and  $p  = 0.15_{-0.09}^{+0.14}$. Adding three more GRBs to the sample, each with a pair of equally possible angle values, can strengthen the anti-correlation to $\rho=-0.46_{-0.08}^{+0.09}$ and $p=0.05_{-0.03}^{+0.07}$. We show that these results are incompatible with non-stationary jets having Gaussian profiles, and that $\gtrsim\!100$ GRBs with observed afterglows would be needed to confirm the potential existence of the angle-variability anti-correlation with $3\sigma$ significance. If the connection is real, GRB jet angles would be constrainable from prompt gamma light curves, without the need of afterglow observations. \end{abstract} 

\begin{keywords}
gamma-ray burst: general -- jets
\end{keywords}



\section{Introduction}
The assumption that long GRB outflows are beamed in collimated jets is evidenced by \emph{jet breaks}, i.e. achromatic breaks observed in afterglow light curves \citep{Kumar2015}. GRB prompt light curves are highly variable \citep{Strong74}, for which one proposed explanation is the angular non-stationarity of jets \citep[see e.g.][]{Roland1994,Zwart1999}, although a more conventional explanation relates it to the time-variability of the emission described by the internal shock model \citep[for details, see e.g.][]{Rees94}. The time of the achromatic break relative to the trigger time of the prompt gamma emission (the so-called \textit{jet break time}, $t_\mathrm{b}$) is often used to deduce an angle value,~$\theta$ \citep[for details, see e.g.][]{Wang18}, which, depending on the assumed luminosity density profile of the jet, is interpreted either as the half opening angle of the jet (uniform profile, see e.g. \citealt{Granot2007}) or the viewing angle between the line-of-sight and the jet axis (power-law or Gaussian profile, see \citealt{Rossi2002} and \citealt{Zhang2002})\footnote{
Note that in \citet{Budai19}, we used the symbol $\theta_\mathrm{c}$ to denote the same quantity we denote by $\theta$ in this article.}.

In \citet{Budai19} we argued that if the dominant source of gamma light curve variabilities is the angular non-stationarity of jets, a statistical connection (correlation or anti-correlation, depending on the jet profile) is expected between $\theta$ and the variability measure of the light curve, $\mathcal{V}$. In this follow-up article we present the results of the test we proposed there, run on 19 long GRBs observed by the Swift satellite, with known $\theta$ values published in \citet{Wang18}. Here, however, we use Spearman's rank correlation instead of the Pearson correlation we proposed in \citet{Budai19}, because the former allows searching for a connection between $\theta$ and $\mathcal{V}$ without assuming that their relationship is linear (see \citealt{Feigelson12} as a reference).

The paper is organised as follows. In Section~\ref{sec:Methods}, we describe the GRB sample and the analysis methods we used. We present the results
of our test in Section~\ref{sec:Results}. In Section~\ref{sec:Conclusion}, we offer conclusions.

\section{Data and methods}\label{sec:Methods}
In \citet{Budai19} we discussed that detector noise and measurement errors can affect the potential detection of a $\theta-\mathcal{V}$ connection. To minimise these effects and to avoid mixing unknown systematics of multiple detectors and data reductions, we chose single-detector samples of GRB light curves and $\theta$ angles, all reduced with the same methods. We took $\theta$ values and their associated errors from the angle catalogue published in \citet{Wang18}. Although more recent catalogues with more $\theta$ values are also available \citep[see e.g.][]{NewTheta}, \citet{Wang18} provides the most sophisticated modelling of the circumburst medium, and applies a stricter selection method of reliable $\theta$ values, thus it is a more conservative choice for our analysis.

We produced mask-weighted, background subtracted Swift BAT light curves ($F$~photon rates and associated errors as a function of time) using the \textit{batgrbproduct}, \textit{batmaskwtevt}, and \textit{batbinevt} tasks \citep{Heasoft} in the 15-150 keV energy band with 10 ms long time bins. We chose the 10 ms bin length because this is the smallest variability time scale of observed GRB light curves \citep[see e.g.][]{Kumar2015}, and also, this was the time step we chose in the simulation we described in \citet{Budai19}. Our final sample of GRBs contains 16 GRBs with single $\theta$ values, and three additional GRBs (GRB~071003, GRB~080319B, GRB~080413B) with pairs of equally possible angle values derived from two observed breaks in their afterglow light curves \citep[see][for details]{Wang18}.

We calculated the $\mathcal{V}$ variability measure of each GRB light curve using the formula we introduced in \citet{Budai19}\footnote{
The codes we used to produce the results of \citet{Budai19} and this paper can be accessed at \url{https://github.com/BMetod}}:
\begin{equation}
 \mathcal{V} = -\frac{1}{T_{90} F_\mathrm{max}} \left[ F_\mathrm{max}-\sum_{i=1}^{N_\mathrm{bin}-1} \left(F_{i}-F_{i+1} \right) \times H\! \left( F_{i}-F_{i+1} \right)  \right]
 \label{eq:var}
\end{equation}
where $F_i$ is the photon rate measured in the $i$th time bin, $F_\mathrm{max}$ is the maximum value of the light curve,  $H(\cdot)$ is the Heaviside step function, and $T_\mathrm{90}$ - the time interval between the epochs when 5~per~cent and 95~per~cent of the total fluence of a GRB is registered by the detector - was taken from the third Swift BAT GRB catalogue~\citep{swift16}.
For each GRB, we created 10~000 light curve realisations by randomising $F_i$ values from normal distributions with means and standard deviations equal to the measured $F_i$ values and errors, respectively. We then calculated $\mathcal{V}$ for each of these realisations using Eq.(\ref{eq:var}). In Table \ref{tab:table1} we report the medians and 68~per~cent confidence limits of the $\mathcal{V}$ distributions, together with the measured $z$ redshift, $T_\mathrm{90}$, and $\theta$ values of the GRBs.

For each of the 16 GRBs with well-defined angle measurements, we randomised 10~000 $\theta$ values from normal distributions with means and standard deviations equal to the measured GRB $\theta$ values and errors (see Table~\ref{tab:table1}), respectively. When this process produced a negative $\theta$, we repeated it until it gave a positive value. We associated the randomised $\theta$ values to the 10~000 $\mathcal{V}$ realisations, thereby creating 10~000 samples of 16~$\theta-\mathcal{V}$ pairs to measure Spearman's $\rho$ and $p$ values (see \citealt{Feigelson12}) with. In Section~\ref{sec:Results} we will report on the medians and 68~per~cent confidence limits we obtained for the resulting $\rho$ and $p$ distributions. Using the three additional GRBs with pairs of equally valid angle measurements, we repeated the same analysis with 19~GRBs eight additional times, one for each combination of all valid $\theta$ values and errors (see Table~\ref{tab:table1}). In Section \ref{sec:Results}, we report on the results of these analyses as well.    

\begin{figure}
 	\includegraphics[width=\columnwidth]{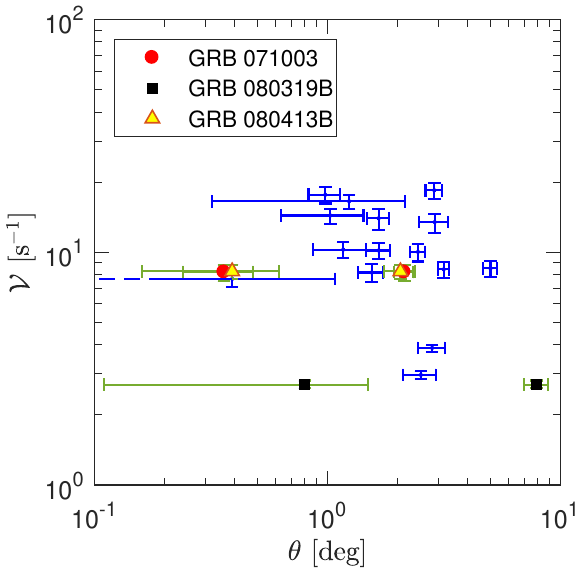}
     \caption{Our GRB sample on the $\theta-\mathcal{V}$ plane based on data in Table \ref{tab:table1}. Light green bars with the circle, square, and triangle markers correspond to the three GRBs with pairs of equally valid $\theta$ measurements. The dashed horizontal line corresponds to GRB~080413A, for which the lower error bar of $\theta$ reaches zero.}
     \label{fig:fig1}
\end{figure}

\section{Results}\label{sec:Results}
Figure~\ref{fig:fig1} shows our GRB sample on the $\theta - \mathcal{V}$ plane, based on Table~\ref{tab:table1}. For the 16~GRBs with single $\theta$ values, the medians and 68~per~cent confidence limits we obtained for the $\rho$ and $p$ distributions are $\rho  = -0.38_{-0.1}^{+0.1}$ and  $p  = 0.15_{-0.09}^{+0.14}$. The fact that 98~per~cent of the 10~000 $\rho$ values we obtained are negative suggests that an anti-correlation may exist between $\theta$ and $\mathcal{V}$, however because of the low number of GRBs in the sample, resulting with relatively high values of $p$, we cannot reject the null hypothesis that $\theta$ and $\mathcal{V}$ have no statistical connection at all.

Table~\ref{tab:table2} shows the medians and 68~per~cent confidence limits of the $\rho$ and $p$ distributions for the eight additional analyses of the 19~GRB samples. For all angle combinations the median $\rho$\nobreakdash-s are negative, with the lowest being $\rho=-0.46^{+0.09}_{-0.08}$, corresponding to the case when for all three GRBs the higher $\theta$ values are assumed. 

\begin{figure}
    \centering
 	\includegraphics[width=.8\columnwidth]{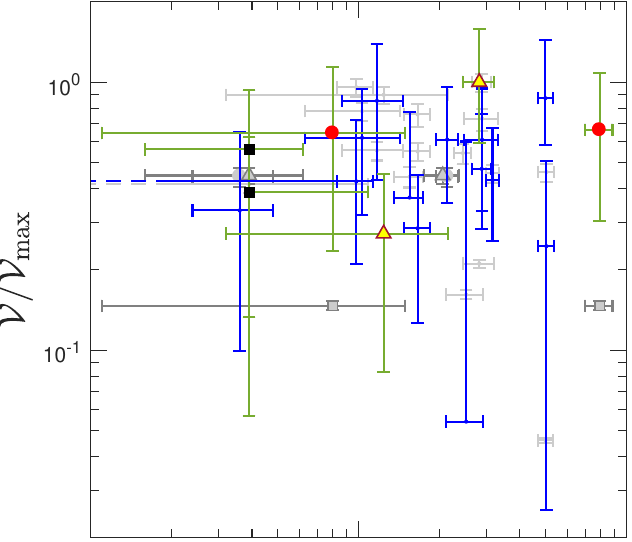}
 	\includegraphics[width=.8\columnwidth]{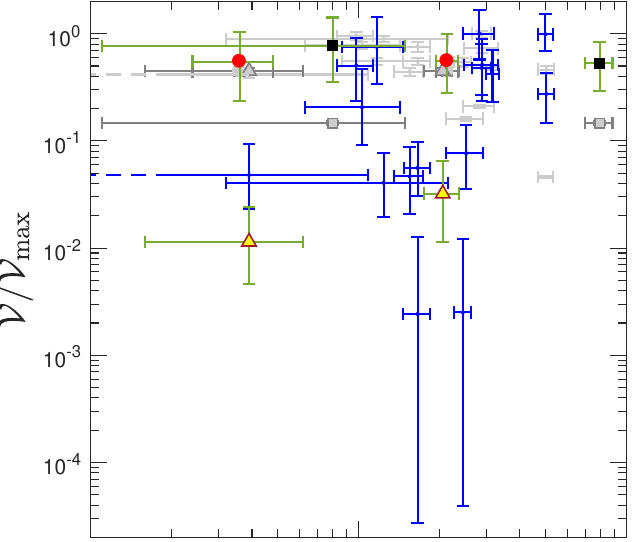}
 	\includegraphics[width=.825\columnwidth]{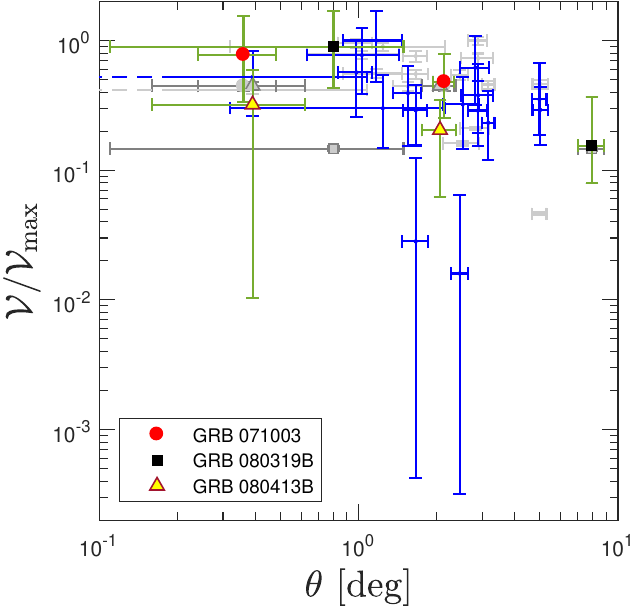}
     \caption{Comparing the simulated samples and the data points of the actual GRB sample plotted in Figure~\ref{fig:fig1}. The luminosity density profiles we used in the simulations are the uniform profile (\emph{upper panel}), the Gaussian profile (\emph{middle panel}) and the power-law profile (\emph{lower panel}). The blue bars show the medians and the 68~per~cent confidence limits of the distribution of the simulated $\mathcal{V}$ values, the green bars with the circle, square, and triangle markers correspond to the three simulated GRBs with pairs of equally valid $\theta$ measurements, the light grey bars show the data points of Figure~\ref{fig:fig1}. The dashed horizontal line corresponds to GRB~080413A, for which the lower error bar of $\theta$ reaches zero. The $\mathcal{V}$ values are normalised with the corresponding  maximal median $\mathcal{V}$ value ($\mathcal{V}_\mathrm{max}$) of each sample.}
     \label{fig:fig2}
\end{figure}

In \citet{Budai19} we used a toy model to simulate light curves of long GRBs with non-stationary jets. We assumed that jets undergo a Brownian random angular motion with a linear restoring force, and showed that although this assumption affects the $\theta-\mathcal{V}$ relationship, the existence of their connection is robust to the choice of the angular motion model. Assuming that all GRBs in our sample have non-stationary jets, we reran the simulation described in \citet{Budai19}, and produced samples of 16 GRB light curves with $z$ and $\theta$ fixed to the values given in Table~\ref{tab:table1} for GRBs with single $\theta$ measurements. 
We used the corresponding $T_\mathrm{90}$ values (see Table~\ref{tab:table1}) as 90~per~cent of the total durations of our simulated light curves, and recalculated the $T_\mathrm{90}$ values of the simulated light curves for the $\mathcal{V}$ calculation using Eq.~\ref{eq:var}.
We produced 10~000 such samples for each of the three possible luminosity density profiles. Figure \ref{fig:fig2} shows the simulated $\mathcal{V}$ values as a function of $\theta$. The data points and error bars represent the medians and the 68~per~cent confidence limits of the distributions of 10~000 $\mathcal{V}$ values for each burst. For each jet profiles the values are normalised with the largest median $\mathcal{V}$ value of the sample for visualisation purposes. In producing Figure \ref{fig:fig2}, we applied the same normalisation on the observed sample as well. The normalisation does not change Spearman's rank correlation however it makes the simulated and observed samples more comparable with each other in the figure. We measured Spearman's $\rho$ and $p$ values for all of the simulated samples. We have found that $\rho < 0$ in 90~per~cent, 34~per~cent, and 0.3~per~cent of the cases for the power-law, the uniform, and the Gaussian profiles, respectively. The $\rho < -0.38$ with $p < 0.15$ condition was satisfied in 28~per~cent, 0.6~per~cent, and zero per~cent of the cases for the three profiles, respectively. These results show, within the limitations of our simulation described in \citet{Budai19}, that the $\rho$ and $p$ values we obtained for our real 16 GRB sample are incompatible with the assumption that GRBs with non-stationary jets have Gaussian luminosity density profiles. 

Using the power-law and uniform profiles only, we also simulated two times 10~000 samples of $N$ GRB light curves with randomly chosen $z$, $T_\mathrm{90}$, and $\theta$ values \citep[see details in][]{Budai19}. We have found that the median $p$ is less than 0.003 if $N>86$ when the power-law, and $N>93$ when the uniform profile is assumed. Thus we conclude that $\gtrsim100$ GRBs with properly measured gamma light curves and $\theta$ values would be needed to detect the $\theta-\mathcal{V}$ connection related to the non-stationarity of jets with a $3\sigma$  significance.

\begin{table}
 	\centering
 	\caption{Measured parameters of GRBs in our sample. The $z$ redshift and $\theta$ values are from \citet{Wang18}, and the $T_\mathrm{90}$ values are from the third Swift BAT GRB catalogue \citep{swift16}. Superscripts \emph{low} and \emph{high} after names of GRBs with pairs of equally valid $\theta$ measurements, denote the lower and higher of these $\theta$ values, respectively. The method we used to measure the $\mathcal{V}$~variabilities of prompt GRB light curves is described in Section \ref{sec:Methods} of this paper.}
 	\label{tab:table1}
 	\begin{tabular}{lcccr} 
 		\hline
GRB name& $z$ & $T_\mathrm{90}\ [\mathrm{s}]$ & $\theta \ [\mathrm{deg}]$ & $\mathcal{V}\ [\mathrm{s}^{-1}]$ \\ 		\hline
GRB 050525A	& $0.606	$ & $			8.84	$ & $		2.52	\pm			0.4		$ & $			2.97_{-0.11}^{+0.11} $ \\[5pt]
GRB 050820A	& $2.6147	$ & $			240.77	$ & $		3.16	\pm			0.18	$ & $			77.67_{-6.39}^{+5.61}$ \\[5pt]
GRB 050922C	& $2.2		$ & $			4.55	$ & $		1.66	\pm			0.19	$ & $			10.19_{-0.79}^{+0.76} $ \\[5pt]
GRB 051109A	& $2.346	$ & $			37.2	$ & $		1.66	\pm			0.18	$ & $			13.98_{-1.50}^{+1.37} $ \\[5pt]
GRB 051111	& $1.55		$ & $		64			$ & $	1.03		\pm		0.4			$ & $		14.37_{-1.14}^{+1.01}    $ \\[5pt]
GRB 060206	& $4.0479	$ & $			7.55	$ & $		2.45	\pm			0.18	$ & $			10.02_{-0.95}^{+0.84} $ \\[5pt]
GRB 060418	& $1.5		$ & $			109.08	$ & $		2.89	\pm			0.41	$ & $			13.49_{-1.39}^{+1.16} $ \\[5pt]
GRB 061126	& $1.5		$ & $			52.62	$ & $		5.02	\pm			0.34	$ & $			8.91_{-0.72}^{+0.67} $ \\[5pt]
$\textrm{GRB 071003}^\mathrm{low}	$& $1.6044	$ & $			148.39	$ & $		0.36	\pm			0.12	$ & $			8.22_{-0.70}^{+0.63} $ \\[5pt]
$\textrm{GRB 071003}^\mathrm{high}	$& $1.6044	$ & $			148.39	$ & $		2.14	\pm			0.2		$ & $			8.22_{-0.70}^{+0.63} $ \\[5pt]
$\textrm{GRB 080319B}^\mathrm{low}	$& $0.937	$ & $			124.86	$ & $		0.8		\pm			0.69	$ & $			2.70_{-0.10}^{+0.11} $ \\[5pt]
$\textrm{GRB 080319B}^\mathrm{high}	$& $0.937	$ & $			124.86	$ & $		7.93	\pm			0.93	$ & $			2.70_{-0.10}^{+0.11} $ \\[5pt]
GRB 080413A	 & $2.433	$ & $			46.36	$ & $		0.39	\pm			0.69	$ & $			7.68_{-0.57}^{+0.52} $ \\[5pt]
$\textrm{GRB 080413B}^\mathrm{low}	$& $1.1		$ & $			8		$ & $		0.39	\pm			0.23	$ & $		8.28_{-0.61}^{+0.56} $ \\[5pt]
$\textrm{GRB 080413B}^\mathrm{high}	$& $1.1		$ & $			8		$ & $		2.06	\pm			0.31	$ & $			8.28_{-0.61}^{+0.56} $ \\[5pt]
GRB 081203A	& $2.1		$ & $			223		$ & $		0.98	\pm			0.15	$ & $			17.66_{-1.52}^{+1.33} $ \\[5pt]
GRB 090618	& $0.54		$ & $		113.34		$ & $	2.82		\pm		0.37		$ & $		3.88_{-0.14}^{+0.11}     $ \\[5pt]
GRB 091029	& $2.752	$ & $			39.18	$ & $		1.24	\pm			0.92	$ & $			16.54_{-1.24}^{+1.09} $ \\[5pt]
GRB 091127	& $0.49		$ & $		6.96		$ & $	1.55		\pm		0.19		$ & $		7.67_{-0.69}^{+0.65}     $ \\[5pt]
GRB 110205A	& $2.22		$ & $		249.42		$ & $	2.88		\pm		0.24		$ & $		17.26_{-1.40}^{+1.21}     $ \\[5pt]
GRB 120729A	& $0.8		$ & $			93.93	$ & $		1.17	\pm			0.3		$ & $			10.28_{-0.84}^{+0.77} $ \\[5pt]
GRB 130427A	& $0.34		$ & $		240.33		$ & $	4.98		\pm		0.32		$ & $		1.26_{-0.03}^{+0.03}     $ \\[5pt]
 		\hline
 	\end{tabular}
 \end{table}

\begin{table}
 	\centering
 	\caption{Results of the Spearman correlation test described in Section~\ref{sec:Methods} run on all 19~GRBs of our sample, including three GRBs (GRB~071003; GRB~080319B; GRB~080414B) that have pairs of equally valid $\theta$ measurements. Rows of the table correspond to samples including different combinations of the lower and the higher $\theta$ values (denoted by \emph{low} and \emph{high}, respectively; see Table~\ref{tab:table1}) of these GRBs. The second and third columns show the medians with the 68~per~cent confidence limits of empirical $\rho$ and $p$ distributions we obtained by correlating 10~000 randomised samples of 19~$\theta-\mathcal{V}$ pairs.}
 	\label{tab:table2}
 	\begin{tabular}{ccc} 
 		\hline
$\theta$ combination & $\rho$ & p \\
 		\hline
(\emph{low; low; low}) & $-0.12_{-0.11}^{+0.10}$ & $0.62_{-0.27}^{+0.26}$ \\[5pt]
(\emph{low; low; high}) & $-0.18_{-0.10}^{+0.11}$ & $0.47_{-0.23}^{+0.28}$ \\[5pt]
(\emph{low; high; low}) & $-0.32_{-0.09}^{+0.08}$ & $0.18_{-0.09}^{+0.14}$ \\[5pt]
(\emph{low; high; high}) & $-0.38_{-0.09}^{+0.08}$ & $0.11_{-0.07}^{+0.10}$ \\[5pt]
(\emph{high; low; low}) & $-0.18_{-0.11}^{+0.10}$ & $0.45_{-0.22}^{+0.29}$ \\[5pt]
(\emph{high; low; high}) & $-0.24_{-0.10}^{+0.10}$ & $0.32_{-0.17}^{+0.25}$ \\[5pt]
(\emph{high; high; low}) & $-0.39_{-0.09}^{+0.08}$ & $0.10_{-0.06}^{+0.10}$ \\[5pt]
(\emph{high; high; high}) & $-0.46_{-0.08}^{+0.09}$ & $0.05_{-0.03}^{+0.07}$ \\[5pt]
 		\hline
 	\end{tabular}
 \end{table}

\section{Conclusions}\label{sec:Conclusion}
Although our tests remained inconclusive on the question whether variabilities of long GRB prompt gamma light curves are dominated by angular non-stationarities of jets, we can already conclude that if this is the case, then it is very unlikely that these jets have a Gaussian luminosity density profile. Based on our simulations, ${\gtrsim\!100}$ long GRBs with gamma light curve and $\theta$ measurements would be needed to statistically confirm the existence of these angular non-stationarities with $3\sigma$ significance.

The sample size is currently limited by the number of $\theta$ measurements. This number will grow in the near future since the extended operation of Swift \citep{swift16} and future missions like SVOM \citep[see e.g.][]{svom11} and CAMELOT \citep[see e.g.][]{Werner18} will add at least tens of new GRBs to our current sample. These existing and future missions can reach a sample size large enough in the upcoming years to carry out the test described in this paper in a decisive way, if they dedicate resources to observing the achromatic breaks as part of their mission goals.

\section*{Acknowledgements}
We would like to thank Rafael de Souza and P\'{e}ter Veres for their helpful insights.

\section*{Data availability}
The data underlying this article are available in GitHub, at \url{https://github.com/BMetod}


\bibliographystyle{mnras}
\bibliography{GRBib}


\bsp	
\label{lastpage}
\end{document}